# Electrodynamics acceleration of electrical dipoles

S. N. Dolya

*Joint Institute for Nuclear Research, Joliot Curie str. 6, Dubna, Russia, 141980*

**Abstract**

This article considers the acceleration of electric dipoles consisting of thin metal plates and dielectric (barium titanate). The dipoles are of a cylindrical shape with a diameter of the cylinder $d_{out}$ = 2 cm and length $d_d$ = 1 sm. Capacity of the parallel-plate capacitor is: C = 278 pF, and it is charged up to the voltage of U = 280 kV. Pre-acceleration of the electric dipoles till velocity $V_{in}$ = 1 km /s is reached by the gas-dynamic method. The finite acceleration is produced in a spiral waveguide, where the pulse is travelling with voltage amplitude $U_{acc}$= 700 kV and power P = 125 MW. This pulse travels via the spiral waveguide and accelerates the injected electric dipoles in the longitudinal direction till the finite velocity $V_{fin}$ = 8.5 km / s over length $L_{acc}$ = 0.77 km.

**Introduction**

There is a known [1] method of accelerating the magnetic dipoles which enables one to accelerate the magnetic dipoles by the running current pulse. To increase the specific magnetic moment, inside the dipoles we place a superconducting coil with the excited current in it.

However, this method of accelerating the magnetic dipoles has a serious drawback. During all the period of acceleration and time of flight to the target, it is necessary to keep the low temperature and superconductivity inside the magnetic dipole. Otherwise, due to a large energy release in the magnetic dipole, it can just collapse.

Another known [2] method of accelerating the charged bodies is as follows: the body is pre-accelerated up to the velocity corresponding to the velocity of injection into the spiral waveguide, then the body is irradiated with an electron beam injected from the electron accelerator, it is electrically charged and finally accelerated by the same voltage pulse running inside the coils of the spiral waveguide.

To set several electrons onto the body is not a problem, but further on when there are many electrons on the body, they will begin to run away from it due to the auto electron emission. Let the field strength for electron emission be as follows: E = 3 * 10$^7$ V / cm. After reaching this field strength, no matter how many electrons you put on the body, they will flow away from the body due to the Coulomb repulsion.

Having planted enough electrons, to plant more, it is necessary to overcome the repulsion of those which are already there. This means that the energy of the electrons, which we want to put onto the body, should be large enough so that they can overcome this Coulomb barrier, reach the body and stay on it.

Coulomb barrier grows for the particles of the cylindrical form linearly with increasing of the diameter and for particles with diameter d = 20 mm it will reach 30 MeV. To overcome it, it will be necessary to accelerate electrons in a specialized accelerator.

The essence of this proposal is to consider the acceleration of the electric dipoles but not the charged bodies.

The acceleration rate for this case has been found to be equal to the following: $\Delta W / \Delta z = (N_e / A)\, e * (2\pi d_d / \lambda_s) * E_{0zw} * \sin\varphi_s$, where $\Delta W / \Delta z$ - a set of plate-capacitor energy per unit of the length, $eN_e / A$ - electrical charge of any sign, located on a plate capacitor and per nucleon in it, $d_d$ - the distance between the plates of a parallel-plate capacitor, $\lambda_s$ - slow down wavelength in the spiral waveguide, $E_{0zw}$ - the amplitude of the wave, $\varphi_s$ - synchronous phase. Acceleration of the plate capacitor is produced in the dielectric channel, which prevents the turn of the dipole by 180 degrees and its deviation from the axis of the acceleration.

**1. Acceleration of electric dipoles**

On the electric dipoles it is possible to set a large number of related charges, i.e., the charges having opposite signs, and, thus, we get a plate capacitor. The total electric charge in the capacitor will be equal to zero, but such electric dipoles will possess a rather large electric dipole moment which can interact with the electric field gradient.

*1. 1.Parameters of the electric dipole*

Let us consider the acceleration of the cylindrical capacitor having an outer diameter $d_{out}$ = 2 cm, the distance between the electrodes is equal to: $d_d$ = 1 cm, where as dielectric we use ceramic capacitor T-900, with a relative permittivity $\varepsilon = 10^3$ and a breakdown voltage U = 28 kV / mm [3], page 321. We assume the density of the ceramic capacitor to be equal to $\rho = 6$ g/cm$^3$ [4].

To find the specific electric charge of the capacitor, we neglect the mass of the metal plates. The square and volume of the dielectric are: $S_d = \pi d_{out}^2 / 4 = 3.14$ cm$^2$, $V_d = S_d * d_d = 3.14$ cm$^3$, respectively, and the mass is: $m_d = V_d * \rho = 18.8$ g.

We find an electric charge on the capacitor. The capacity of the flat capacitor in the practical system of coordinates is as follows:

$$C = \varepsilon\varepsilon_0 S_d/d_d = 278 \text{ pF}, \qquad (1)$$

where $\varepsilon_0 = 8.85 * 10^{-12}$ F / m – dielectric permittivity of vacuum. Such a capacitor can be charged up to 280 kV [3], page 321, so that the electric charge (expressed in practical units), will be:

$$Q = CU = 7.8 * 10^{-5} \text{ Coulomb}, \qquad (2)$$

i.e., the capacitor will contain: $N_e = 7.8 * 10^{-5} * 6 * 10^{18} = 4.7 * 10^{14}$ electrons.

The number of nucleons in this cylinder is equal to: $A = 1.13 * 10^{25}$ nucleons, the ratio of the charge to the mass in such a capacitor will be equal to the following: $(N_e / A) e = 4.1 * 10^{-11}$.

*1.2. Acceleration of the dipoles*

Similarly to the gradient of the magnetic field accelerating the magnetic dipole, the acceleration of the electric dipole is carried out by the gradient of the electric field of the wave:

$$F_e = (N_e / A) e * dE_{zw} / dz. \qquad (3)$$

As for the magnetic dipole, the gradient of the electric field of the wave is $dE_{zw} / dz = k_3 * E_{zw}$, $k_3 = 2\pi/\lambda_s$, where $\lambda_s = \lambda_0 * \beta_{ph}$ - slowdown wave length in the structure, $\beta_{ph} = V_{ph} / c$ - relative phase velocity of the wave in a spiral waveguide, $c = 3 * 10^{10}$ cm / s - velocity of light in vacuum.

Finally,

$$F_e = (N_e / A) e * (2\pi d_d/\lambda_s) * E_{zw0} * \sin\varphi_s, \qquad (4)$$

where $E_{zw0}$ - the amplitude of the electric field strength on the axis of the spiral.

Before substituting the numbers into formula (4), we make a few general remarks.

First of all, it should be stressed that force Fe accelerating the electric dipoles does not depend on the length of the dipole ($d_d$). Indeed, while decreasing $d_d$, the capacitance C of a the plate capacitor grows, but at the same time the breakdown voltage U decreases, thus, the charge stored in the capacitor does not depend on parameter $d_d$.

When parameter $d_d$ decreases the mass of the dielectric reduces and the specific electric charge - the charge per unit of the mass, increases. However, while decreasing $d_d$, the accelerating force $F_e$ acting on the dipole, decreases. This is due to the fact that the accelerating force is the difference between the repulsive force of one pole and the electric force of the other pole. From (4) it is clear that this force is greater, the greater the distance $d_d$ is between the dipole poles.

Moreover, the accelerating force acting on the electric dipole is independent of $S_d$ – the transverse dipole square. Indeed, with the growth of this square the charge stored in the capacitor increases, but simultaneously, in the same proportions, the mass of the dielectric also grows. This leads to the independence of the specific electrical charge of the capacitor on its transverse cross section. It is important to mention that in the case with a magnetic dipole the situation is different. The magnetic moment of the coil with the current grows as the square of the coil, i.e. the square of the increasing radius. The mass of the coil increases as the perimeter of the coil, i.e., linearly with the increasing radius, that results in linear increasing of the specific magnetic moment with the increasing radius of the coil with the current.

Now we substitute the numbers into the formula (4). The first factor ($N_e$ / A)e determines the maximum electric charge per nucleon, which can be stored in the capacity. As it is shown above, this charge is determined only by the properties of the substance, in this case, - by the properties of the capacitor ceramic, i.e. - by the highest possible relative permittivity $\varepsilon$ and maximum breakdown voltage U. Probably, materials with better properties will be developed later.

The second factor $2\pi d_d/\lambda s$ is determined by the ratio of the length of the dipole $d_d$ to the slowdown wavelength $\lambda_s$ in the spiral waveguide. At a large length of the accelerator it will be necessary to divide the accelerator into separate sections each of them will be individually supplied with power. Then it will be possible to accelerate the dipoles in each section at the optimal frequency for each section, which in the case of the spiral waveguide is defined by the following ratio: $\lambda_s = 2\pi r_0$, where $r_0$ - the radius of the spiral. The diameter of the spiral may be chosen slightly larger than $d_{out}$ - diameter of the capacity, for example: $d_{out}/2r_0 = 0.5$, and then the parameter is equal to $d_d/r_0 = 0.5$.

Finally, for $E_{zw} = E_{zw0} * \sin\varphi_s = 250$ kV / cm, we find:

$$F_e = (N_e / A) e * (2\pi d_d/\lambda_s) E_{zw} = 5.2 * 10^{-4} \text{ eV} / (m * \text{nucleon}), \quad (5)$$

thus, to achieve the finite velocity $V_{fin} = 8.5$ km / s,
$W_{fin} = 0.4$ eV / nucleon, the acceleration length will be required to be equal to:

$$L_{acc} = W_{fin} / F_e = 0.77 \text{ km}. \quad (6)$$

**2. The structure of the accelerator**

Fig. 1 shows the scheme of the accelerator.

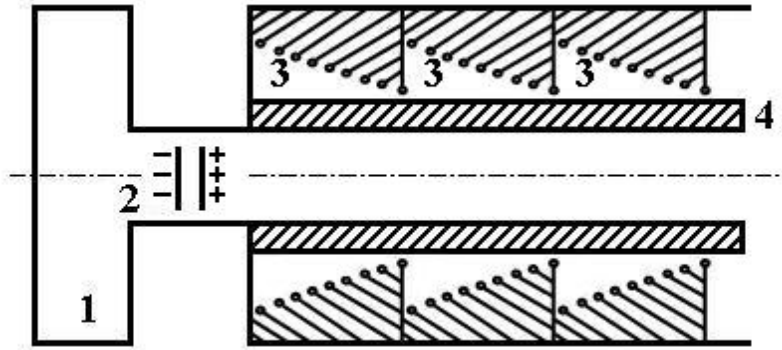

Fig. 1

Fig. 1 shows: 1 - gun, carrying out the gas-dynamic acceleration of electric dipoles, 2 - electric dipoles, pre-charged electrical capacitors, 3 - section of the spiral waveguide, 4 – dielectric channel located on the axis of the system where the acceleration of the electric dipoles is carried out.

*2.1. Pre-acceleration of the electric dipoles by using the gas-dynamic method*

To accelerate the electric dipoles by the field of the traveling wave, this wave must be very slow. It should be mentioned that the relative velocity $\beta = 10^{-6}$ corresponds to the normal velocity equal to: $V = 0.3$ km / s and is less than the velocity of the sound in the air. The gas-dynamic acceleration method does not allow one to achieve a higher velocity than $V_g = 2$ km / s.

For example, specifications of the gun AP 35/1000, produced by the German company "Rheinmetall" are as follows: the initial rate of shooting $V_{in} = 1.5$ km / s, the diameter of the projectile: $d_{sh} = 35$ mm. The company "Mauser" is developing an aircraft gun with a caliber (diameter of the projectile) $d_{sh} = 30 - 35$ mm and a projectile velocity $V_{in} = 1.8$ km / s.

We take the initial velocity of the electric dipoles to be reached after the gas-dynamic acceleration, equal to: $V_{in} = 1$ km / s.

*2. 2. Spiral step*

We have to take a very small step of spiral winding since we have chosen the radius of the spiral to be equal to: $r_0 = 2$ cm, then to get the spiral slowdown equal to $\beta_{ph\ in} = 3.3 * 10^{-6}$, where $\beta_{ph\ in} = V_{in} / c$ - initial phase velocity, expressed in the units of the light velocity and coinciding with the initial velocity of the electric dipoles.

The slowdown in the spiral is purely geometrical. In the simplest case the phase velocity of the wave, expressed in the units of the light velocity, in the spiral waveguide is equal to:

$$\beta_{ph} = \operatorname{tg} \Psi, \qquad (7)$$

where tg $\Psi$ - the tangent of the winding angle, the tangent, in the case of large decelerations, is equal to: tg $\Psi = h/2\pi r_0$ –the ratio of the step of spiral winding to the perimeter of the coil.

Besides purely geometric slowdown of the wave in the spiral, it is possible to slow it down additionally if to place the spiral totally inside the dielectric medium with a relative permittivity $\varepsilon$. For barium titanate near the Curie point, it is possible to achieve the following values: $\varepsilon = 8 * 10^3$, [5], page 557.

But we take a smaller value of ε: ε = 1.28 * 10³, the relationship between the phase velocity of the wave propagating in the spiral and its parameters in this case, can be written in the following form [6],

$$\beta_{ph} = \text{tg } \Psi / \varepsilon^{1/2}, \qquad (8)$$

The area inside the spiral should be left free of the dielectric because the acceleration of the electric dipoles will take place in this region along the axis of the spiral. Then, for the spiral where the dielectric has filled the region located between the coil and the external screen, the dispersion equation – the equation relating the parameters of the spiral with the phase velocity of the wave looks as follows, [6],

$$\beta_{ph} = \sqrt{2} * \text{tg } \Psi / \varepsilon^{1/2}. \qquad (9)$$

For the start of the spiral, where the velocity of the electric dipoles is equal to $V_{in}$ = 1 km / s, $\beta_{ph\ in}$ = 3.3 * 10⁻⁶, $r_0$ = 2 cm, ε = 1.28 * 10³, from (9) we find that the winding step of the spiral must be equal to:

$$h = 10^{-3} \text{ cm}. \qquad (10)$$

The amplitude of the field strength $E_{zw0}$, which we have chosen, is equal to: $E_{zw0} = E_{zw} / \sin\varphi_s$ = 350 kV / cm, $\sin\varphi_s$ = 0.7.

At step h = 10⁻³ cm = 10 μ there is a risk of the electric breakdown of the dielectric. Breakdown voltage of the polyimide is 300 MV / m, [5], page 550, or 300 V / μ, thus, you can choose the spiral structure to be as follows: copper coil with a cross section of 8 μ and isolation of polyimide 2microns thick.

*2. 3. The required wave power*

Relationship between the power flux and the wave strength on the spiral is given by the following formula [6],

$$P = (c/8)E_{zw0}^2 r_0^2 \beta_{ph}\{(1+I_0K_1/I_1K_0)(I_1^2 - I_0I_2) + \varepsilon(I_0/K_0)^2(1+I_1K_0/I_0K_1)(K_0K_2 - K_1^2)\}. \qquad (11)$$

The argument of the modified Bessel functions of the first and second types presented in the curly brackets is the value of x = 2πr₀/λs, which we have chosen to be equal to: x = 1. Then, for this argument the second term in curly brackets is much greater than the first term, and the curly bracket itself is equal to: $\{\} = 3.77 * \varepsilon$. Substituting the numbers into the formula (11), $\varepsilon = 1280$, we find

$P (W) = 3 * 10^{10} * 3.5 * 3.5 * 10^{10} * 4 * 3.3 * 10^{-6} * 3.77 * 1.28 * 10^{3} /$
$/ (8 * 300 * 300 * 10^{7}) = 125$ MW.

In order to reach the field strength at the axis of the spiral to be equal to $E_{zw0} = 350$ kV / cm, it is required to introduce power P = 125 MW.

This power can be achieved by the pulse technology.

We expand the sinusoidal pulse [6], corresponding to the half-wave $E_{pulse} = E_{0pulse} \sin (2\pi/T_0) t$, $2\pi/T_0 = \omega_0$, $\omega_0 = 2\pi f_0$ in a Fourier row:

$$f_1 (\omega) = (2 / \pi)^{1/2} \int_0^{T_0/2} \sin\omega_0 t * \sin(\omega t) dt. \quad (12)$$

The pulse spectrum is rather narrow and covers the frequency range from 0 to $2\omega_0$. Since the spiral waveguide dispersion (in dependence of the phase velocity on frequency) is weak, it can be expected that the full range of frequencies from 0 to $2\omega_0$ will propagate approximately with the same phase velocity.

As a result, the half-wave sinusoidal pulse in vacuum will spread out only due to increasing of the phase velocity of the wave. In this case matching of the spiral waveguide with a power feeder is necessary to carry out in the frequency band: $\Delta f \approx \omega_0/2\pi$.

We introduce the notion of the pulse amplitude $U_{acc}$, related with the field strength at the axis of the spiral $E_{zw0}$ by the following ratio [6]:

$$U_{acc} = E_{zw0} \lambda_s / 2\pi, \lambda_s = \beta \lambda_0, \lambda_0 = c/f_0. \quad (13)$$

In this case, the vacuum wavelength $\lambda_0$ is: $\lambda_0 = \lambda_s / \beta_{in} = 4.18 * 10^{6}$ cm, the wave frequency: $f_0 = c/f_0 = 7.1 * 10^{3}$ Hz, half-life - the duration of the pulse on the basis is equal to: $T_0 / 2 = 1/2f_0 = 90$ µs.

Thus, the amplitude of the voltage pulse propagating along the spiral must be equal to: $U_{acc} = E_{zw0} * \lambda_s/2\pi = 700$ kV. The Table below summarizes the main parameters of the accelerator.

Table. Parameters of the accelerator.

| Parameter | Value |
|---|---|
| Number of electrons per nucleon in the electric dipole, (Z/A)e | $(Z/A)e = 4.1*10^{-11}$ |
| The ratio of the dipole to the slowdown-wave length $2\pi d_d /\lambda_s$ | $2\pi d_d /\lambda_s = 0.5$ |
| Wave power in Watts, P | P = 125 MW |
| The speed of the electric dipoles, the initial - final, $\beta = \beta_{ph}$ | $\beta_{ph} = 3.3*10^{-6} - 2.83*10^{-5}$ |
| The radius of the spiral, $r_0$ | $r_0 = 2$ cm |
| The frequency of the wave, $f_0$, | $f_0 = 7.1*10^3$ Hz |
| The tension of the electric. field $E_{zw0}$ | $E_{zw0} = 350$ kV/cm |
| The length of the accelerator, $L_{acc}$ | $L_{acc} = 0.77$ km |
| Pulse duration, $\tau$ | $\tau = 90$ µs |
| The amplitude of the voltage, $U_{acc}$ | $U_{acc} = 700$ kV |

*2. 4. Phase stability*

It is known that in the traveling wave the phase stability region is on the front slope of the wave pulse. Indeed, if the particle is faster than the wave, it will get into the weakening field, and, finally, the speeding up pulse will soon catch up with the particle.

If the particle is behind its phase, it will get into the region of the strengthening field and acquire more energy in comparison with what would have been obtained in the synchronous phase and, eventually, the particle will catch up with its phase.

Thus, from the viewpoint of the mutual position of the particle and the acceleration pulse there is only possible case when the pulse «pushes» but not «pulls» the particle. In our calculations we have chosen the synchronous phase to be equal to: $\varphi_s = 45^0$, $\sin \varphi_s = 0.7$. To achieve a greater rate of acceleration, it is possible to choose a greater value $\varphi_s = 60^0$, $\sin \varphi_s = 0.87$, but then at the acceleration it will be needed to meet more strict requirements.

It is known that while acceleration of particles in azimuthal - symmetrical field, the phase stability corresponds to the radial instability. This means that when we push the particle by the pulse, we push it not only forward but also a bit on the radius. The force acting on the particle increases linearly with the growth of particle deviation from the axis and the initial deviation increases exponentially.

To keep the particles near the axis, it is necessary to introduce focusing while accelerating, i.e., it is necessary to use additional forces, fields which will return the particle to the axis of the acceleration.

*2. 5. Preventing the turn of the dipole by $180^0$ in the electric field of the pulse and keeping it on the axis of the dipole acceleration*

In the acceleration of the dipoles there is a new problem which did not exist while accelerating the point particles. The pulse accelerating the dipole will lead to a roll-over of the dipole - turn it by 180 degrees.

This problem is easier to see with the example of the acceleration of the magnetic dipoles and conventional magnets by the running current pulse. If such a pulse, pushing the magnet, to substitute by another magnet, it can be seen that the magnet being pushed will not be easy to push from the same sign pole of the pushing magnet but it will try to turn by 180 degrees and pull itself to the opposite-sign pole.

The action of the pair of forces leading to the "reversal" of the dipole is summed up in comparison with the "difference" forces accelerating the dipole and leading to the radial displacement of the center of mass.

The simplest solution that prevents reversal of the magnetic dipoles by 180 degrees in the accelerating wave field is the imposition of the uniform external magnetic field. It won't influence the acceleration of the dipole because of its homogeneity but will only hold the dipole against the reversal in the space.

By analogy with the uniform magnetic field which does not affect the acceleration of the dipoles but keeps the magnetic dipoles against the reversal by 180 degrees, it is possible to use the uniform electric field to hold the electric dipole against the reversal.

The intensity of the electric field must be at least of a larger wave amplitude, i.e. $E_{keep}$ > 350 kV / cm. When acceleration length $L_{acc}$ is ≈ 1 km, the electrostatic field for this purpose is not acceptable since in this case a tremendous difference of potentials would take place.

The uniform electric field can be formed by the induction method. Let the length of the inductor be (along the acceleration axis) $l_{ind}$ = 10 cm. Such a length will be covered by the electric dipole with velocity $V_{in}$ = 1 km / s during the period of time equal to $\tau_{ind}$ = $10^{-4}$ s, hence, the induction method should create tension $U_{ind} = E_{keep} * l_{ind}$ = 3.5 MV for the period of time $\tau_{ind}$ = $10^{-4}$ s.

Multiplying these values, we find that the required amount of the magnetic flux is F = 350 T * $m^2$. Let this flux be created by iron re-magnetization from the value of -1 T till the value of +1 T, then the cross-section of the iron in the inductor should be $S_{ind}$ = 175 $m^2$. Then at the length of the inductor along the acceleration axis - $l_{ind}$ = 10 cm, its radial length should be: $r_{ind}$ = 1750 m, that is not justified either.

It is possible to hold the electric dipoles near the acceleration and at the same time - to keep them against turning by 180 degrees, if you keep the acceleration of the electric dipoles in a narrow dielectric channel, located on the axis of the system. In this case at the longitudinal acceleration of the electric dipoles in the channel the dipole will undergo friction with the walls of the channel.

*2. 6. The influence of friction*

We consider how strong the friction of the side surface of the electric dipoles over the internal surface of the channel, will influence the acceleration.

The side surface of the electric dipoles must be dielectric to it avoid the short current in the capacitor. The channel must be also dielectric not to screen the electric field accelerating the electric dipoles along the spiral structure.

Actually, the friction force directed against the accelerating force is equal to the coefficient of friction multiplied by the radial force pressing the electric dipoles towards the inner surface of the channel. In its turn, the radial force proportional to the acceleration force and deviation of the electric dipole from the axis is absent when the center of the electric dipoles is located on the axis of the system.

Let the deviation of the electric dipoles from the axis of the system is 1% of its radius, $d_{out} / 2 = 1$ cm, i.e. $\Delta r = 100$ μ. From the expansion of the Bessel function of the first order with a small argument, it is clear that a pair of forces acting on a dipole is equal to the following:

$$F_r = (2\pi\Delta r/\lambda_s) (N_e / A) e * (2\pi d_d/\lambda_s) E_{zw}. \qquad (14)$$

A typical value of the coefficient of friction, for example, by cellophane [5], page 128, is: $k_{fr} = 0.4$. On the one hand, the coefficient of friction increases when the friction surfaces are in vacuum, on the other hand, it decreases with increasing of the relative velocity of motion. We assume that for our velocity of the electric dipoles moving within the channel, the friction coefficient is about the same value. Then, the ratio of force directed against velocity, $F_f$, to the force of acceleration is equal to

$$F_f / F_e = k_{fr} * (2\pi\Delta r/\lambda_s) = 2 * 10^{-2}, \qquad (15)$$

that will be compensated by phase stability.

According to Hooke's law we find the elastic force $F_{elas}$, which will prevent the compression of the electric dipoles:

$$\Delta r / d_{out} = F_{elas} / (E * S_c), \qquad (16)$$

where: $\Delta r / d_{out} = 5 * 10^{-3}$ - relative compression, E - Young's modulus, $S_c$ - the contact surface.

At a height of segment $\Delta r$, the length of chord a can be found from the formula:

$$a = 2 (\Delta r * d_{out} - \Delta r^2)^{1/2} \approx 2 (\Delta r * d_{out})^{1/2}, \qquad (17)$$

and, if the contact surface is: $S_c = d_d * a = 3 * 10^{-5}$ m². A typical value of Young's modulus E for plastics is equal to $E = 10^8$ N/m², [3], p.53, and from (16) we find that the value of force $F_{elas}$ is: $F_{elas} = 15$ N, that is comparable with radial force Fr which must be multiplied by the number of nucleons $A = 1.13 * 10^{25}$, $F_r = 5*10^{-3}*5.2*10^{-4}*1.6*10^{-19}*1.13*10^{25} = 5$ N.

It is clear that the elastic force returning the center of gravity of the electric dipoles onto the axis of the system will prevent a significant radial displacement of the center of gravity of the electric dipoles relatively the axis of the system.

**Conclusions**

The efficiency of the acceleration of charged cylindrical bodies decreases with increasing of the diameter of the cylinder. This is due to the fact that the electric charge is located on the cylinder surface. The square of the unit of the cylinder length increases as the radius of the cylinder, and the volume and mass increase as the square of the radius. As a result, the charge per unit of the (nucleon) mass reduces with increasing of the radius as 1 / r. From a certain radius the acceleration rate of the electric dipoles by the wave field gradient becomes greater than the acceleration rate of the charged bodies.

The choice of electric dipoles in the form of a plate capacitor allows the electric dipoles to reach the first order space velocity at a distance less than one kilometer.